\documentclass[english]{article}
\usepackage[T1]{fontenc}
\usepackage[latin9]{luainputenc}
\usepackage{geometry}
\usepackage{amstext}
\usepackage{amssymb}
\usepackage{feyn}
\usepackage{tikz}
\usepackage{graphicx}
\usepackage{array}
\usepackage{booktabs}

\usepackage{amsmath}
\usepackage{amsfonts}

\usepackage[normalem]{ulem}
\usepackage{color}
\usepackage{listings,braket}
\usepackage{caption}
\usepackage{subcaption}
\usepackage{float}
\definecolor{darkgreen}{rgb}{0,0.35,0}
\definecolor{Rood}{rgb}{1, 0, 0}

\makeatletter
\usepackage[english]{babel}
\usepackage{feyn}

\makeatother

\begin{document}

\title{\textbf{On the Feynman path integral formulation of the Bell-Clauser-Horne-Shimony-Holt inequality in Quantum Field Theory}}

{\author{\textbf{G.~Peruzzo$^1$}\thanks{gperuzzofisica@gmail.com},
		\textbf{S.~P.~Sorella$^2$}\thanks{silvio.sorella@gmail.com},\\\\\
		\textit{{\small $^1$Instituto de F\'{i}sica, Universidade Federal Fluminense,
				}}\\
		\textit{{\small Campus da Praia Vermelha, Av. Litor\^{a}nea s/n, }}\\
		\textit{{\small 24210-346, Niter\'{o}i, RJ, Brasil}}\\
		\textit{{\small $^2$UERJ -- Universidade do Estado do Rio de Janeiro,}}\\
		\textit{{\small Instituto de F\'{\i}sica -- Departamento de F\'{\i}sica Te\'orica -- Rua S\~ao Francisco Xavier 524,}}\\
		\textit{{\small 20550-013, Maracan\~a, Rio de Janeiro, Brasil}}\\
	}

\date{}

\maketitle
\begin{abstract}
By employing a free scalar Quantum Field Theory model previously introduced \cite{Peruzzo:2022pwv}, we attempt at formulating the Bell-CHSH inequality within the Feynman path integral. This possibility relies on the observation that the Bell-CHSH inequality exhibits a natural extension to Quantum Field Theory in such a way that it is compatible with the time ordering $T$. By treating the Feynman propagator as a distribution and by introducing a suitable localizing set of compact support smooth test functions, we work out the path integral setup for the Bell-CHSH inequality, recovering the same results of the canonical quantization. 
\end{abstract}

\section{Introduction} 
The study of the Bell-Clauser-Horne-Shimony-Holt \cite{Bell:1964kc,Bell:1964fg,Clauser:1969ny,Freedman:1972zza,Clauser:1974tg,Clauser:1978ng}} inequality is one of the cornerstones of the physics of the entanglement, as documented by the large literature available on the subject  in Quantum Mechanics. \\\\From the theoretical point of view of relativistic Quantum Field Theory, it seems fair to state that the study of the Bell-CHSH inequality is yet to be considered at its beginning. Let us mention that the topic has a great phenomenological  interest in view of the  future experiments at LHC, see \cite{Ashby-Pickering:2022umy} and refs. therein. \\\\The study of the Bell-CHSH inequality in Quantum Field Theory started with the pioneering work of  \cite{Summers:1987fn,Summ,Summers:1987ze,Summers:1988ux,Summers:1995mv} who, making use of Algebraic Quantum Field Theory \cite{Haag:1992hx}, showed that even free fields lead to a violation of the Bell-CHSH inequality. This  result highlights the strength of entanglement in Quantum Field Theory \cite{Verch}. Though, many aspects remain still to be unraveled. Let us quote, for example, the general treatment of interacting Quantum Field Theories as well as the construction of a $BRST$ invariant setup for the Bell-CHSH inequality in Abelian and non-Abelian gauge theories. \\\\More recently, following \cite{Summers:1987fn,Summ,Summers:1987ze,Summers:1988ux,Summers:1995mv}, we have constructed an explicit Quantum Field Theory model built out by means of a massive free scalar field and of a suitable Bell-CHSH operator exhibiting  a violation of the Bell-CHSH inequality at the quantum level \cite{Peruzzo:2022pwv}. The results obtained in \cite{Peruzzo:2022pwv} relied on the use of the canonical quantization. \\\\The aim of the present work is that of pursuing the investigation of the Bell-CHSH inequality in Quantum Field Theory. More precisely, we shall attempt at formulating the Bell-CHSH inequality within the framework of the Feynman path integral, a topic which, to our knowledge, has not yet been addressed. Needless to say, the path integral formulation will enable us to study the Bell-CHSH inequality for interacting field theories by employing the dictionary of the Feynman diagrams, including the BRST invariant formulation of Abelian and non-Abelian gauge theories. \\\\Several issues arise when trying to achieve the path integral formulation of the Bell-CHSH inequality. Willing to present them briefly, we might start to mention that the Feynman path integral is intrinsically related to the chronological time ordering $T$. A second issue concerns the complex character of the Feynman propagator, {\it i.e.} 
\begin{equation}
\Delta_F(x-x') = \int \frac{d^4p}{(2\pi)^4} \frac{e^{-ip(x-x')}}{p^2-m^2 +i\varepsilon}  \neq (\Delta_F(x-x'))^{\dagger} \;. \label{feynm} 
\end{equation}
Both aspects have to be properly addressed when comparing the  Hermitian expression of the Bell-CHSH correlator obtained via canonical quantization    with the corresponding quantum correlator evaluated with the Feynman path integral. \\\\As we shall see in the following, these issues can be faced by making use of smeared fields, namely
\begin{equation} 
\varphi(f) = \int_\Omega d^4x \;\varphi(x) f(x) \;, \label{cs}
\end{equation} 
where $f(x)$ is a test function with compact support,  $\Omega$, belonging to the  space ${\cal C}_{0}^{\infty}(\mathbb{R}^4)$, {\it i.e.} to the space of smooth infinitely differentiable functions decreasing as well as their derivatives faster than any power of $(x) \in \mathbb{R}^4$ in any direction \cite{Haag:1992hx}.  As it is apparent from \eqref{cs}, the introduction of the test function $f(x)$ has the effect of localizing the field $\varphi(x)$ in the region $\Omega$. Working with smeared fields has many advantages. First, the use of a suitable set of test functions will enable us to introduce a Bell-CHSH correlator compatible with the time ordering $T$, a basic requirement in order to have a path integral formulation. Secondly, the analytic properties of the Fourier transform of test functions belonging to ${\cal C}_{0}^{\infty}(\mathbb{R}^4)$, see \cite{Gelfand}, 
allow to handle the Feynman $i\varepsilon$ prescription by the usual Cauchy theorem, so as to recover exactly the result of the canonical setup.  \\\\Moreover, besides the pure mathematical aspects related to the introduction of the test functions, we would like to point out  that, in the case of the study of the violation of the Bell-CHSH inequality, the smearing procedure acquires a rather clear and simple physical meaning. Looking  at the details of one of the most recent experiments \cite{zgiu1}, one realizes that issues like the so-called causality loophole, {\it i.e.} the effective experimental implementation of the space-like separation  between the two polirazers, namely Alice and Bob's devices, is very carefully handled.  Both polarizers are randomly rotating while the pair of entangled photons emitted by the source is flying towards them, so that it turns out to be impossible for the photons to communicate each other about the direction in which their respective polarization is being measured. This very sophisticated setup has the practical effect of closing the causality loophole.  Willing thus to achieve a relativistic Quantum Field Theory framework for the Bell-CHSH inequality, it seems to us very helpful  employing  a clear localization procedure which stays as close as possible with the experiments. This is precisely the role played by the introduction of the test functions, namely: allowing for a well defined localization procedure in space-time.  \\\\The paper is organized as follows. In Section \eqref{BCHSH} we elaborate on the comparison between the Bell-CHSH operator of Quantum Mechanics and that of Quantum Field Theory, pointing out the very basic requirement of compatibility with the time ordering $T$. This section contains a detailed construction of the Quantum Field Theory operators entering the Bell-CHSH correlator. Here, we shall rely on a field theory model built out with a pair of free scalar fields. As we shall see, this model enable us to make use of a squeezed state, allowing for the maximal violation of the Bell-CHSH inequality. Section \eqref{Feynman} is devoted to the Feynman path integral formulation of the Bell-CHSH inequality by establishing the equivalence with the canonical formalism. In Section \eqref{conclus} we collect our conclusion. \\\\Overall, for the benefit of the reader, we attempted at presenting the various topics in a self-contained way.

\section{The Bell-CHSH inequality in Quantum Mechanics and in relativistic Quantum Field Theory: smearing and compatibility with the time ordering $T$}\label{BCHSH}

\subsection{The Bell-CHSH inequality in Quantum Mechanics: a short reminder}\label{reminder} 
Let us begin by reminding the construction of the Bell-CHSH operator in Quantum Mechanics,  as presented in textbooks, see for example \cite{Nielsen,Peres,Zwiebach}. One starts by introducing a two spin $1/2$ operator
\begin{equation}
{\cal C}_{CHSH} = \left[ \left( {\vec \alpha} \cdot {\vec\sigma}_A    + {\vec \alpha'} \cdot {\vec\sigma}_A  \right) \otimes  {\vec \beta}\cdot{\vec \sigma}_B + \left( {\vec \alpha} \cdot {\vec \sigma}_A    - {\vec \alpha'} \cdot {\vec\sigma}_A  \right) \otimes  {\vec \beta'}\cdot{\vec \sigma}_B \right] \;, \label{chshqm}
\end{equation}
where $(A,B)$ refer to Alice and Bob, $\vec \sigma$ are the spin $1/2$ Pauli matrices and $({\vec \alpha},{\vec \alpha'},{\vec \beta},{\vec \beta'})$ are four arbitrary unit vectors.\footnote{Due to $\sigma_i \sigma_j = \delta_{ij} + i\varepsilon_{ijk}\sigma_k$, it follows that $({\vec n} \cdot {\vec\sigma})^2 =1$ for any unit vector $|\vec n|=1$. }
The operator \eqref{chshqm} has the renowned form
\begin{equation}
{\cal C}_{CHSH} = (A+A')B + (A-A')B' \;.  \label{AB1}
\end{equation}
with $(A,A')$ and $(B,B')$ denoting the Alice and Bob spin operators
\begin{equation} 
A= {\vec \alpha} \cdot {\vec\sigma}_A \;, \qquad A'={\vec \alpha'} \cdot {\vec\sigma}_A\;, \qquad B={\vec \beta}\cdot{\vec \sigma}_B \;, \qquad B'={\vec \beta'}\cdot{\vec \sigma}_B
\end{equation} 
fulfilling the following commutation relations 
\begin{equation} 
\left[ A,B \right] =0\;, \qquad \left[ A,B' \right] =0\;, \qquad \left[ A',B \right] =0 \;, \qquad \left[ A',B' \right] =0 \;. \label{crts}
\end{equation} 
Moreover, $(A,A')$ and $(B,B')$ are all Hermitian, with eigenvalues ${\pm 1}$. \\\\On the basis of the so-called local realism of hidden variables \cite{Bell:1980wg}, one expects that 
\begin{equation} 
| {\cal C}_{CHSH} | \le 2 \;, \label{in}
\end{equation} 
for any possible choice of the unit vectors $({\vec \alpha},{\vec \alpha'},{\vec \beta},{\vec \beta'})$. \\\\Nevertheless, it turns out that this inequality is  violated by Quantum Mechanics, due to entanglement. In fact, when evaluating the Bell-CHSH correlator in Quantum Mechanics, {\it i.e.} $ \langle \psi | {\cal C}_{CHSH} |\psi \rangle $, where $|\psi\rangle$ is an entangled state as, for example, the Bell singlet, one gets 
\begin{equation}
| \langle \psi | {\cal C}_{CHSH} |\psi \rangle | = 2\sqrt{2} \;, \qquad |\psi\rangle = \frac{ |+\rangle_A\otimes |- \rangle_B - 
|-\rangle_A\otimes |+\rangle_B}{\sqrt{2} }  \label{vb} \;. 
\end{equation} 
The bound $2\sqrt{2}$ is known as Tsirelson's bound \cite{tsi1,tsi2}, providing the maximum violation of the $CHSH$ inequality \eqref{in}. The experiments carried out over the last decades, see \cite{zgiu1} and refs therein, have largely confirmed the violation of the Bell-CHSH inequality, being in very good agreement with the bound $2\sqrt{2}$.

\subsection{Construction of the Bell-CHSH in Quantum field theory: localization and compatibility with the time ordering $T$} \label{T}

\subsubsection{Basic features of the canonical quantization}\label{cq}
In order to address the issue of the construction of the analogue of the Bell-CHSH operator \eqref{chshqm} in Quantum Field Theory, it is useful to recall here a  few basic properties of the canonical quantization of a free massive scalar field \cite{Haag:1992hx}:  
\begin{equation} 
{\cal L} =   \frac{1}{2} \left( \partial^\mu \varphi \partial_\mu \varphi - m^2 \varphi^2 \right) \;.  \label{cnq1}
\end{equation} 
Expanding $\varphi$ in terms of annihiliation and creation operators, one gets 
\begin{equation} 
\varphi(t,{\vec x}) = \int \frac{d^3 {\vec k}}{(2 \pi)^3} \frac{1}{2 \omega(k,m)} \left( e^{-ikx} a_k + e^{ikx} a^{\dagger}_k \right) \;, \qquad k^0= \omega(k,m) = \sqrt{{\vec{k}}^2 + m^2}  \;, \label{qf}
\end{equation} 
where 
\begin{equation} 
[a_k, a^{\dagger}_q] = (2\pi)^3 2\omega(k,m) \delta^3({\vec{k} - \vec{q}}) \;, \qquad [a_k, a_q] = 0\;, \qquad [a^{\dagger}_k, a^{\dagger}_q] =0\;, \label{ccr}
\end{equation}
are the canonical commutation relations. A quick computation shows that
\begin{equation} 
\left[ \varphi(x) , \varphi(y) \right] = i \Delta_{\textrm{PJ}} (x-y) = 0 \; \qquad {\rm for } \; \;\; (x-y)^2<0 \;, \label{caus} 
\end{equation}
where $\Delta_{\textrm{PJ}}(x-y) $ is the Lorentz  invariant causal Pauli-Jordan function, encoding the principle of relativistic causality 
\begin{itemize} 
\item 
\begin{equation}
 \Delta_{\textrm{PJ}}(x-y) = \frac{1}{i} \int \frac{d^4k}{(2\pi)^3} (\theta(k^0) - \theta(-k^0)) \delta(k^2-m^2) e^{-ik(x-y)}\;, \label{it1}
\end{equation} 
\item 
\begin{equation} 
\Delta_{\textrm{PJ}}(x-y) = - \Delta_\textrm{{PJ}}(y-x) \;, \qquad (\partial^2_x + m^2) \Delta_{\textrm{PJ}}(x-y) = 0 \;, \label{it2}
\end{equation} 
\item
\begin{equation} 
\Delta_{\textrm{PJ}}(x-y) = \left( \frac{\theta(x^0-y^0) -\theta(y^0-x^0)}{2\pi} \right) \left( -\delta((x-y)^2)  +m \frac{\theta((x-y)^2)  J_1(m\sqrt{(x-y)^2}) }{ 2 \sqrt{(x-y)^2}}\right) \;,
\end{equation} 
where $J_1$ is the Bessel function. 
\end{itemize} 
It is known that  expression  \eqref{qf} is a too singular object, being in fact an operator valued distribution \cite{Haag:1992hx}. To give a well defined meaning to eq.\eqref{qf}, one introduces the smeared field 
\begin{equation} 
\varphi(h) = \int d^4x \;\varphi(x) h(x) \;, \label{sm} 
\end{equation} 
where $h(x)$ is a test function belonging to the space of compactly supported smooth functions ${\cal C}_{0}^{\infty}(\mathbb{R}^4)$. The support of $h(x)$, $supp_h$, is the region in which the test function $h(x)$ is non-vanishing. Moving to the Fourier space   
\begin{equation}
{\hat h}(p) = \int d^4x \; e^{ipx} h(x)  \;, \label{fft}
\end{equation} 
expression \eqref{sm} becomes 
\begin{equation} 
\varphi(h) = \int \frac{d^3 {\vec k}}{(2 \pi)^3} \frac{1}{2 \omega(k,m)} \left( {\hat h}^{*}(\omega(k,m),{\vec k}) a_k + {\hat h}(\omega(k,m),{\vec k}) a^{\dagger}_k \right)  = a_h + a^{\dagger}_h\;, \label{smft} 
\end{equation}
where $(a_h,a^{\dagger}_h)$ stand for 
\begin{equation} 
a_h = \int \frac{d^3 {\vec k}}{(2 \pi)^3} \frac{1}{2 \omega(k,m)}  {\hat h}^{*}(\omega(k,m),{\vec k}) a_k \;, \qquad 
a^{\dagger}_h = \int \frac{d^3 {\vec k}}{(2 \pi)^3} \frac{1}{2 \omega(k,m)} {\hat h}(\omega(k,m),{\vec k}) a^{\dagger}_k \;. 
\end{equation} 
One sees that the smearing procedure has turned the too singular object $\varphi(x)$, eq.\eqref{qf}, into an operator acting on the Hilbert space of the system, eq.\eqref{smft}. When rewritten in terms of the operators 
$(a_f,a^{\dagger}_g)$, the canonical commutation relations \eqref{ccr} read
\begin{equation} 
\left[ a_h,a^{\dagger}_{h'}\right]  = \langle h | h' \rangle \;, \label{ccrfg}
\end{equation}
where $ \langle h | h' \rangle$ denotes the Lorentz invariant scalar product between the test functions $h$ and $h'$. {\it i.e.} 
\begin{equation} 
\langle h | h' \rangle = \int \frac{d^3 {\vec k}}{(2 \pi)^3} \frac{1}{2 \omega(k,m)}  {\hat h}^{*}(\omega(k,m),{\vec k}) {\hat h'}(\omega(k,m), {\vec k}) = 
\int \frac{d^4 {\vec k}}{(2 \pi)^4} 2\pi \;\theta(k^0) \delta(k^2-m^2)  {\hat h}^{*}(k) {\hat h}'(k)  \;. \label{scpd}
\end{equation} 
The scalar product \eqref{scpd} can be recast in configuration space. Taking the Fourier transform, one has 
\begin{equation} 
\langle h | h' \rangle = \int d^4x d^4x'\; h(x) {\cal D}(x-x') h'(x')  \;, \label{confsp} 
\end{equation} 
where ${\cal D}(x-x')$ is the so-called Wightman function
\begin{equation} 
{\cal D}(x-x') = \langle 0| \varphi(x) \varphi(x') |0 \rangle = \int \frac{d^3 {\vec k}}{(2 \pi)^3} \frac{1}{2 \omega(k,m)} e^{-ik(x-x')}  \;, \qquad k^0=\omega(k,m) \;. \label{Wg}
\end{equation} 
which can be decomposed as 
\begin{equation} 
{\cal D}(x-x') = \frac{i}{2} \Delta_{\textrm{PJ}}(x-x')   + H(x-x') \;, \label{decomp}
\end{equation} 
where $H(x-x')=H(x'-x)$ is the real symmetric quantity \cite{Scharf1} 
\begin{equation} 
H(x-x') = \frac{1}{2} \int \frac{d^3 {\vec k}}{(2 \pi)^3} \frac{1}{2 \omega(k,m)} \left( e^{-ik(x-x')} + e^{ik(x-x')}
 \right) \qquad k^0=\omega(k,m)\;. \label{H}
\end{equation} 
Finally, the commutation relation (\ref{caus}) can be expressed in terms of smeared fields as
\begin{equation} 
\left[ \varphi(h) , \varphi(h') \right] =i  \Delta_{\textrm{PJ}}(h,h') \label{comm_smeared}
\end{equation}
where $h$, $h'$ are test functions and
\begin{equation} 
\Delta_{\textrm{PJ}}(h,h')= \int d^4x \; d^4x' h(x) \Delta_{\textrm{PJ}}(x-x')  h'(x'). \label{pauli_jordan_smeared}
\end{equation}
Therefore, the causality condition in terms of smeared fields becomes
\begin{equation} 
\left[ \varphi(h) , \varphi(h') \right] = 0, \label{caussm}
\end{equation}
if $supp_h$ and $supp_{h'}$ are space-like.

\subsubsection{Weyl operators} \label{Weyl}
For further use, let us present  here the so-called Weyl operators.    The Weyl operators are bounded unitary operators built out by exponentiating the smeared field, namely 
\begin{equation} 
{\cal A}_h = e^{i {\varphi}(h) }\;, \label{Weyl}
\end{equation}
where ${\varphi}(h)$ is the smeared field defined in eq.\eqref{sm}.  Making use of the following relation
\begin{equation}
e^A \; e^B = \; e^{ A+B +\frac{1}{2}[A,B] } \;, \label{exp_AB}
\end{equation} 
valid for two operators $(A,B)$ commuting with $[A,B]$, one immediately checks that the Weyl operators give rise to the following algebraic structure
\begin{eqnarray}
{\cal A}_h \;{\cal A}_h' & =  & e^{- \frac{1}{2} [{\varphi}(h), {\varphi}(h')] }\;{\cal A}_{(h+h')} = e^{ - \frac{i}{2} \Delta_{\textrm{PJ}}(h,h')}\;{\cal A}_{(h+h')}  \nonumber \\
{\cal A}^{\dagger}_h & = & {\cal A}_{(-h)}\;, \label{algebra} 
\end{eqnarray} 
where $\Delta_{\textrm{PJ}}(h,h')$ is defined in eq.\eqref{pauli_jordan_smeared}. Also, using the canonical commutation relations written in the form \eqref{ccrfg}, for the vacuum expectation value of ${\cal A}_h$, one gets 
\begin{equation} 
\langle 0| \; {\cal A}_h \; |0 \rangle = \; e^{-\frac{1}{2} {\lVert h\rVert}^2} \;, \label{vA}
\end{equation} 
where the vacuum state $|0\rangle$ is the Fock vacuum: $a_k|0\rangle=0$ for all modes $k$.

\subsubsection{Algebra of the Bell operator and the time ordering $T$} \label{constr}
We are now ready to face the issue of the construction of the Bell-CHSH operator in Quantum Field Theory. We follow here the setup outlined in \cite{Summers:1987fn,Summ,Summers:1987ze,Summers:1988ux,Summers:1995mv} and introduce the notion of {\it eligibility}. A set of four field operators $(A,A')$ and $(B,B')$ are called eligible for the Bell-CHSH inequality if:
\begin{itemize} 
\item they are all Hermitian 
\begin{equation}
A=A^{\dagger}\;, \qquad A'=A'^{\dagger}\;, \qquad B=B^{\dagger}\;, \qquad B'=B'^{\dagger} \;, \label{herm}
\end{equation}
\item obey the condition
\begin{equation} 
  (A,A')\; {\rm and\;} (B,B')\; {\rm are\; bounded\; operators, \;taking \;values\; in\; the\; interval\;} [-1,1] \;, \label{m1}  
\end{equation} 
\item Alice's operators $(A,A')$ commute with Bob's operators $(B,B')$, namely 
\begin{equation} 
\left[ A,B \right] =0\;, \qquad \left[ A,B' \right] =0\;, \qquad \left[ A',B \right] =0 \;, \qquad \left[ A',B' \right] =0 \;. \label{crtsfield}
\end{equation} 
\end{itemize}
 
\begin{figure}[!ht] 

	\centering
	\includegraphics[scale=0.5]{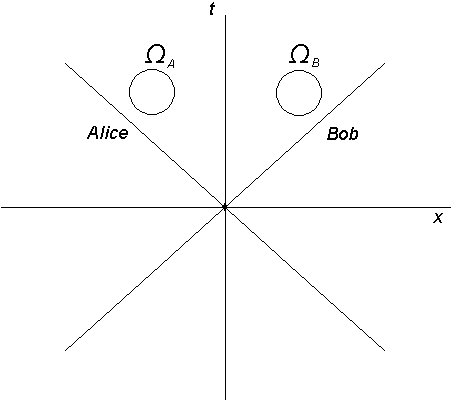}
	\caption{Location of the labs of Alice and Bob in a two-dimensional spacetime diagram.}
	\label{contorno_c}
\end{figure}

 Let us focus now on eq.\eqref{crtsfield}. Its fulfillment requires a well specified localization property of both Alice and Bob operators in space-time. More precisley, relying on the relativistic causality,  eq.\eqref{comm_smeared}, one is led to demand that the supports of Alice's test functions $(f,f')$ belong to a space-time region $\Omega_A$ which is space-like with respect to the region $\Omega_B$ containing the supports of Bob's test functions $(g,g')$, {\it i.e.} 
\begin{equation} 
( supp_{(f,f')})   \;\;\;\;  {\rm space-like\; with \; respect \;to} \;\;\;\; (supp_{(g,g')}) \;, \label{supp} 
\end{equation}
see Fig.\eqref{contorno_c}. 

\noindent These considerations make clear the key role of the relativistic causality when analysing the Bell-CHSH inequality in Quantum Field Theory. The use of smeared fields and of the test functions acquires a clear physical meaning: $(f,f')$ and $(g,g')$ act as space-time localizers for Alice's and Bob's operators, implementing in a practical way the fundamental principle of relativistic causality. \\\\Furthermore, the demand of space-like separation between Alice and Bob has a relevant consequence for the time ordering $T$. In fact, if $O_1(x)$ and $O_2(y)$ are two field operators and $(x-y)^2<0$, it follows that 
\begin{equation}
[O_1(x), O_2(y)] = 0 \;, \qquad (x-y)^2 < 0 \;, \label{o1o2}
\end{equation}
so that the $T$ product reduces to the identity, {\it i.e.}
\begin{equation}
T \left( O_1(x) O_2(y) \right) = \theta(x^0-y^0) O_1(x) O_2(y) + \theta(y^0-x^0) O_2(y) O_1(x) = O_1(x) O_2(y) \;. \label{Tid}
\end{equation}
An immediate consequence of all this is that the Bell-CHSH combination is, by construction, left invariant by the time ordering $T$, namely 
\begin{equation} 
T \left( (A+A')B + (A-A')B') \right) = (A+A')B + (A-A')B'  \;. \label{comp}
\end{equation}
Although equation \eqref{comp} looks a simple consequence of the requirements \eqref{herm},\eqref{m1},\eqref{crtsfield}, it seems fair to state that it expresses a deep property of the Bell-CHSH particular combination. It paves the route for the Feynman path integral formulation. 

\subsection{Example of violation of the Bell-CHSH inequality in free Quantum Field Theory} \label{ex-qft}
In order to provide an explicit example of the violation of the Bell-CHSH inequality in the vacuum state, we shall consider a model consisting of a pair of free massive real scalar fields $(\varphi_A, \varphi_B)$, \begin{equation} 
{\cal L} =   \frac{1}{2} \left( \partial^\mu \varphi_A \partial_\mu \varphi_A - m^2_A \varphi_ A\varphi_A \right)+\frac{1}{2} \left( \partial^\mu \varphi_B \partial_\mu \varphi_B - m^2_B \varphi_B\varphi_B \right) \;. 
\label{clact}
\end{equation}
From the canonical quantization, we have 
\begin{eqnarray} 
\varphi_A(t,{\vec x}) & = & \int \frac{d^3 {\vec k}}{(2 \pi)^3} \frac{1}{2 \omega(k,m_A)} \left( e^{-ikx} a_k + e^{ikx} a^{\dagger}_k \right) \;, \qquad k^0= \omega(k,m_A)  \;, \nonumber \\
\varphi_B(t,{\vec x}) & = & \int \frac{d^3 {\vec k}}{(2 \pi)^3} \frac{1}{2 \omega(k,m_B)} \left( e^{-ikx} b_k + e^{ikx} b^{\dagger}_k \right) \;, \qquad k^0= \omega(k,m_B)  \;,\label{alice_bob_fields}
\end{eqnarray} 
where the only non-vanishing commutators among the annihilation and creation operators are
\begin{eqnarray} 
\left[ a_k, a^{\dagger}_q\right]  & = &(2\pi)^3 2\omega(k,m_A) \delta^3({\vec{k} - \vec{q}}) \, , \nonumber \\
\left[ b_k, b^{\dagger}_q\right]  & = &(2\pi)^3 2\omega(k,m_B) \delta^3({\vec{k} - \vec{q}})\, .
 \label{ccr_alice_bob}
\end{eqnarray}
To have well defined operators in the Fock-Hilbert space, these fields are smeared with test functions, resulting in $\left(\varphi_A(h),\;\varphi_B(h)\right)$. It is thus  straightforward to evaluate the following commutation relations for the smeared fields
\begin{eqnarray} 
	\left[ \varphi_A(h) , \varphi_A(h') \right] &=& i \Delta_{\textrm{PJ}}^{m_A}(h,h')\; , \nonumber \\
	\left[ \varphi_B({\tilde h}) , \varphi_B({\tilde h'}) \right] &=& i  \Delta_{\textrm{PJ}}^{m_B}({\tilde h},{\tilde h}')\;, \nonumber \\
	\left[ \varphi_A(h) , \varphi_B({\tilde h}) \right] &=& 0\;,
		 \label{caussm}
\end{eqnarray}
valid for any pair of test functions $(h,h')$, $({\tilde h}, {\tilde h'})$. The presence of the Pauli-Jordan function $\Delta_{\textrm{PJ}}$ in expressions \eqref{caussm}  implements the relativistic causality  in the model. In fact, if $supp_h$ and $supp_{h'}$ are space-like as well as those of $({\tilde h},{\tilde h'})$, then the commutator of the corresponding  smeared fields vanishes.  The Fock vacuum  of the model  is defined as being the state $| 0 \rangle$ such that
\begin{eqnarray}
a_{k}|0 \rangle=0\,,\qquad
b_{k}|0 \rangle=0\, , \label{fock_vacuum_a_b}
\end{eqnarray}
for any mode $k$. \\\\Let us turn now to the Bell-CHSH inequality. We start  with a general consideration on the algebraic relations fulfilled by the four  operators $(A,A')$, $(B,B')$. \\\\Let us introduce the  Hermitian Bell-CHSH field operator 
\begin{equation} 
 {\cal C}_{CHSH} = (A + A') B + (A-A')B' \;. \label{chshft}
\end{equation}
In agreement with \cite{Summers:1987fn,Summ,Summers:1987ze,Summers:1988ux,Summers:1995mv}, we shall say that the Bell-CHSH inequality is violated at the quantum level in the vacuum if 
\begin{equation} 
|\langle 0 | {\cal C}_{CHSH} | 0 \rangle| >  2 \;. \label{vq}
\end{equation} 
We proceed by rewriting the vacuum expectation value of the Bell-CHSH operator as 
\begin{equation} 
\langle 0 | {\cal C}_{CHSH}  | 0 \rangle  > = \langle 0 | {\cal U}^\dagger \;{\cal U} \;\left( (A + A')\;{\cal U}^\dagger \;{\cal U} \;B + (A-A')\;{\cal U}^\dagger \;{\cal U}\; B' \right) \;{\cal U}^\dagger \;{\cal U}  | 0 \rangle \;, \label{UBCHSH}
\end{equation}  
where $\cal U$ stands for a unitary operator: 
\begin{equation} 
{\cal U}^\dagger {\cal U} = 1 \;. \label{unit}
\end{equation} 
Introducing now the unitary equivalent operators 
\begin{equation} 
{\hat A} = {\cal U}^\dagger A {\cal U}\;, \qquad {\hat A'} = {\cal U}^\dagger A' {\cal U}\;, \qquad {\hat B} = {\cal U}^\dagger B {\cal U}\;, \qquad {\hat B'} = {\cal U}^\dagger B {\cal U}\;, \label{hatop}
\end{equation} 
it is easily checked that $({\hat A},  {\hat A'})$ and $({\hat B},  {\hat B'})$ fulfill the same algebraic relations of 
$({A},  { A'})$ and $({ B},  {B'})$, namely 
\begin{eqnarray} 
\hat{A}^2 & =& 1\;, \qquad \hat{A'}^2 =1\;, \qquad \hat{B}^2 =1 \;, \qquad  \hat{B'}^2=1 \;, \nonumber \\
\left[ {\hat A}, {\hat B} \right] & = & 0 \;, \qquad \left[ {\hat A}, {\hat B'} \right]  =  0 \;, \qquad \left[ {\hat A'}, {\hat B} \right]  =  0 \;, \qquad \left[ {\hat A'}, {\hat B'} \right]  =  0 \;. \label{salg}
\end{eqnarray} 
Suppose now that the unitary operator ${\cal U}$ is such that its action on the vacuum $|0\rangle$ creates a two-mode  entangled state $|\eta \rangle$, namely 
\begin{equation} 
{\cal U} |0\rangle = |\eta \rangle \;. \label{create} 
\end{equation} 
This is the case, for instance, of the two-mode squeezed state. In such a case, the unitary operator is nothing but the squeezed operator \footnote{We point out that the test functions can be always normalized to $1$, namely
\begin{equation}
f\rightarrow \frac{f}{||f||}\Rightarrow ||f||=1\,.
\end{equation}
As a consequence $\left[a_{f},a_{f}^\dagger\right]=1$. Similarly, $\left[b_{g},b_{g}^\dagger\right]=1$. }
\begin{eqnarray} 
{\cal U} & = & e^{\frac{r}{2} \left( a^\dagger_f b^\dagger_g - a_f b_g \right) } \;, \nonumber \\
|\eta \rangle & = & (1-\eta^2)^{\frac{1}{2}} \sum_{n=0} \eta^n | n_f n_g \rangle \;, \qquad \eta = \rm{ tgh} (r) \;, \label{squeezed} 
\end{eqnarray} 
and 
\begin{equation} 
| n_f n_g \rangle = \frac{1}{n} (a^\dagger_f)^n (b^\dagger_g)^n |0\rangle \;. \label{abn} 
\end{equation} 
Therefore, we get the equality
\begin{equation}
\langle 0 | (A+A')B+(A-A')B' |0\rangle = \langle \eta | (\hat{A}+\hat{A}')\hat{B}+(\hat{A}-\hat{A}')\hat{B}' |\eta \rangle \, . \label{vacuum_squeezed}
\end{equation}
This equation states that the vacuum expectation value of the Bell-CHSH operator can be obtained by evaluating the expectation value of the unitary equivalent combination $ (\hat{A}+\hat{A}')\hat{B}+(\hat{A}-\hat{A}')\hat{B}'$ in the squeezed state $|\eta\rangle$. Equation \eqref{vacuum_squeezed} is very helpful, both from theoretical and computational points of view. In fact, following \cite{Summ}, the squeezed state $|\eta\rangle$ can be rewritten as the sum of even and odd modes, \emph{i.e.} 
\begin{equation}
|\eta \rangle = \left(1-\eta^2\right)^{\frac{1}{2}}\left(\sum_{n=0}^{\infty}\eta^{2n}|2n_f 2n_g\rangle +\sum_{n=0}^{\infty}\eta^{2n+1}|(2n_{f}+1)(2n_{g}+1)\rangle \right)\,.
\end{equation}
One defines the operators $\hat{A}_i=\left(\hat{A},\hat{A}'\right)$ and $\hat{B}_k=\left(\hat{B},\hat{B}'\right)$ as \cite{Summ}:
\begin{eqnarray}
\hat{A}_{i}|2n_f\,\cdot \rangle &=& e^{i \alpha_i} | \left(2n_f+1\right)\,\cdot \rangle\,, \nonumber \\
\hat{A}_{i}| \left(2n_f+1\right)\,\cdot \rangle&=& e^{-i \alpha_i} |2n_f\,\cdot \rangle,
\end{eqnarray}
and
\begin{eqnarray}
\hat{B}_{k}|\cdot \, 2n_g \rangle &=& e^{i \beta_k} |\cdot \, \left(2n_g+1\right) \rangle\,, \nonumber \\
\hat{B}_{k}|\cdot \, \left(2n_g+1\right) \rangle&=& e^{-i \beta_i} |\cdot \, 2n_g \rangle,
\end{eqnarray}
where $\left(\alpha_i,\,\beta_k\right)$ are arbitrary real quantities. The operators $\hat{A}_i$ act only on the first entry, while the operators $B_k$ only on the second.\\\\ From a quick computation, it turns out that
\begin{equation}
\langle \eta | (\hat{A}+\hat{A}')\hat{B}+(\hat{A}-\hat{A}')\hat{B}' |\eta \rangle = \frac{2\eta}{1+\eta^2}\left[\cos\left(\alpha_1+\beta_1\right)+\cos \left(\alpha_2+\beta_1\right)+\cos\left(\alpha_1+\beta_2\right)-\cos\left(\alpha_2+\beta_2\right)\right].
\end{equation}
Setting 
\begin{equation}
\alpha_1=0,\qquad \alpha_2= \frac{\pi}{2}, \qquad \beta_1=-\frac{\pi}{4}, \qquad \beta_2=\frac{\pi}{4}\,,
\end{equation}
one gets
\begin{equation}
\langle 0 | {\cal{C}}_{CHSH} | 0 \rangle = \frac{2 \eta}{1+\eta^2}2\sqrt{2}\,,
\end{equation}
which attains Tsirelson bound for $\eta \approx 1$,
\begin{equation}
\langle 0 | {\cal{C}}_{CHSH} | 0 \rangle \approx 2\sqrt{2}\,.
\end{equation}
This result is in full agreement with that of \cite{Summ}.

\section{Feynman path integral formulation of the Bell-CHSH inequality}\label{Feynman} 
Relying on the results and observations of the previous sections, let us discuss now the Feynman path integral formulation of the Bell-CHSH inequality. To that end, let us start by introducing the  generating functional  
${\cal Z}(j)$ of the time ordered correlation functions:
\begin{equation} 
{\cal Z}(j) = \frac{\int [D\varphi]\; e^{i \left( S(\varphi) + \int d^4x j\varphi \right)}}{\int [D\varphi]\; e^{i \left( S(\varphi)  \right)}} = e^{-\frac{i}{2} \int d^4x d^4y j(x) \Delta_F(x-y) j(y) } \;, \label{F1} 
\end{equation} 
where $\Delta_F(x-y)$ is the Feynman propagator, {\it i.e.}
\begin{equation}
\Delta_F(x-y) = \int \frac{d^4p}{(2\pi)^4} \frac{e^{-ip(x-y)}}{p^2-m^2 +i\varepsilon}  \;. \label{F2} 
\end{equation}
It is useful to remind here the expression of $\Delta_F(x-y)$ in configuration space. Using the same notations of 
\cite{Greiner}, $\Delta_F(x-y)$ can be written as 
\begin{equation} 
\Delta_F(x-y) = \frac{1}{2}(\theta(x^0-y^0)- \theta(y^0-x^0)) \Delta_{PJ}(x-y) - i H(x-y) \;,  \label{F3}
\end{equation} 
where $\Delta_{PJ}$ is the Pauli-Jordan function, eq.\eqref{it1}, and $H$ is the symmetric expression of eq.\eqref{H}. Explicitly: 
\begin{eqnarray} 
\Delta_F(x-y) &=& - \frac{1}{4\pi} \delta((x-y)^2) + \frac{ m \theta((x-y)^2) }{8\pi \sqrt{(x-y)^2} }\left( J_1(m \sqrt{(x-y)^2} ) - i N_1 (m \sqrt{(x-y)^2} ) \right) \nonumber \\
& - & \frac{im \theta(-(x-y)^2) }{4\pi^2 \sqrt{-(x-y)^2 }} K_1(m \sqrt{-(x-y)^2 })  \;, \label{F4}
\end{eqnarray}
where, from the second line, one observes the well known non-causal behavior of the Feynman propagator. In the above expression, $J_1$ is the Bessel function, $N_1$ the Neumann function and $K_1$ the modified Bessel function. \\\\To achieve the equivalence between the path integral and the canonical formalism we evaluate, for example, the correlation functions of two Weyl operators in both cases, the aim being that of showing that 
\begin{equation} 
\langle e^{i \varphi(h)}\; e^{i \varphi(h')} \rangle_{Feyn} = \langle 0| e^{i \varphi(h)}\; e^{i \varphi(h')} |0\rangle_{can} \;, \label{canequiv}
\end{equation}
where the supports of the two test functions $(h,h') \in {\cal C}_{0}^{\infty}(\mathbb{R}^4)$ are taken as located in the positive half-plane $t>0$, as in Fig.\eqref{contorno_c},  and are space-like, {\it i.e.} 
\begin{equation} 
( supp_{(h)})   \;\;\;\;  {\rm space-like\; with \; respect \;to} \;\;\;\; (supp_{(h')}) \;, \label{supphh} 
\end{equation}
a feature which, as already underlined, reduces the chronological ordering $T$ to unity. \\\\The computation of the left hand side of eq.\eqref{canequiv} is easily done with the help of \eqref{F1}, namely 
 \begin{equation} 
{\cal Z}(j) = \langle e^{i \varphi(j)} \rangle_{Feyn} \;. \label{F5} 
\end{equation}
Therefore, setting 
\begin{equation} 
j = h+ h' \;, \label{jhh}
\end{equation}
it follows that 
\begin{equation} 
\langle e^{i \varphi(h)}\; e^{i \varphi(h')} \rangle_{Feyn} = e^{-\frac{i}{2}  \Delta_F(h+h',h+h') } = 
e^{-\frac{i}{2} \left( \Delta_F(h,h)+ 2\Delta_F(h,h')+ \Delta_F(h',h') \right)}\;, \label{F6} 
\end{equation} 
 where $(\Delta_F(h,h), \Delta_F(h,h'),  \Delta_F(h',h'))$ denote the smeared expressions 
 \begin{eqnarray} 
 \Delta_F(h,h') & = & \int d^4x d^4y\; h(x) \Delta_F(x-y) h'(y)   \;, \label{smear1} \\
 \Delta_F(h,h) & =& \int d^4x d^4y\; h(x) \Delta_F(x-y) h(y) \;, \qquad \Delta_F(h',h')  = \int d^4x d^4y\; h'(x) \Delta_F(x-y) h'(y). \;. \label{smear2}
 \end{eqnarray} 
 One sees that eq.\eqref{F6} demands the evaluation of two kinds of smeared expressions involving the Feynman propagator. Let us first consider expression \eqref{smear1}, where the smearing is done with respect to two different test functions: $(h,h')$. Reminding that the supports of $h$ and $h'$ are space-like, 
 eq.\eqref{supphh}, one can rely directly on expression  \eqref{F3}, from which one  realizes that the Pauli-Jordan term $\Delta_{PJ}(x-y)$ does not contribute since it vanishes for space-like separations. Thus, 
 \begin{equation}
  \Delta_F(h,h') = - i H(h,h') = - i \int d^4x d^4y\; h(x)  H(x-y) h'(y) \;. \label{Hhh}
 \end{equation} 
 Though, owing to the general definition of the scalar product of test functions in terms of Wightman two point function, eqs.\eqref {confsp},\eqref{Wg},\eqref{decomp}, it follows that 
 \begin{equation} 
  \Delta_F(h,h') = -i \langle h | h' \rangle \;. \label{dfhh}
 \end{equation}
 Let us now focus on the expressions $\Delta_F(h,h)$ and $\Delta_F(h',h')$ in eq.\eqref{smear2}. These quantities require a different handling, as the Feynman propagator is smeared exactly over the same support. We proceed by moving to the Fourier space, {\it i.e.}
 \begin{equation} 
  \Delta_F(h,h) = \int \frac{d^4p}{(2\pi)^4} \; \frac{1}{p^2-m^2 +i\varepsilon} | h(p_0, \vec{p})|^2 \;, \label{Ft1}
 \end{equation} 
 where $h(p_0, \vec{p})$ is the Fourier transform of $h(x)$ 
 \begin{equation} 
 h(p_0, \vec{p}) = \int d^x\; e^{i px} h(x) \;. \label{ft2}
 \end{equation}
 As is well known \cite{Gelfand}, being $h(x)$ a Schwartz type test function, its Fourier transform displays an exponential decay at larhe |p|. Moreover, since $h(x) \in {\cal C}_{0}^{\infty}(\mathbb{R}^4)$ and its support is located in the positive half-plane $t>0$, Fig.\eqref{contorno_c}, it follows that $h(p_0, \vec{p})$ can be analytically continued to an entire function in the complex $p_0$ plane, decaying very fast for large values of $Im(p_0)$ in the positive complex $p_0$ half-plane. \\\\Let us illustrate this relevant property with a simple one-dimensional example taken from  Chapter II of \cite{Gelfand}. Consider the function $f(x) \in {\cal C}_{0}^{\infty}(\mathbb{R})$ defined as 
 \begin{equation}
f(x) = 
     \begin{cases}
       {\cal C} e^{-\frac{1}{(x-a)^2(x-b)^2}} &\quad \text{if} \;  x\in [a,b] \;, a,b>0  \\
       0 &\quad\text{if} \;  x \notin [a,b]  \;,  \label{bf}
     \end{cases} 
    \end{equation}
where ${\cal C}$  is a normalization factor.  The function $f(x)$ is a smooth function, infinitely differentiable, which  is non-vanishing only in the interval $x\in [a,b]$. For the Fourier transform, we have 
\begin{equation} 
{\hat f}(p) = \int_{-\infty}^{\infty}  dx \; e^{ipx} f (x)  \;, \label{f}
\end{equation} 
Moreover, since $f(x)$ has compact support, the integral \eqref{f} becomes 
 \begin{equation} 
{\hat f}(p) = \int_{a}^{b}  dx \; e^{ipx} f (x)  \;, \label{fh}
\end{equation} 
 and can be analytically continued to an entire function in the complex plane 
 \begin{equation} 
 z= p + i \tau\;,  \qquad {\hat f}(z) = \int_{a}^{b}  dx \; e^{ipx} e^{-\tau x} f (x)  \;, \label{fh1}
 \end{equation} 
 Notice that the analytic continuation of ${\hat f}(p)$ to an entire function is possible thanks to the fact that $f(x)$ has compact support, so that the integral \eqref{fh1} does exist for all values of $\tau$. Of corse $ {\hat f}(z)$ decays very fast to zero when $Im(z)=\tau \rightarrow \infty$: 
 \begin{equation} 
 \lim_{\tau \rightarrow \infty} {\hat f}(p+i\tau) = 0 \;. \label{lim}
 \end{equation} 
 Therefore, as a consequence of these properties, we can evaluate expression  \eqref{Ft1} by employing the residue Cauchy theorem in the complex $p_0$ plane, by closing the contour to infinity in the upper positive imaginary half plane, getting nothing but 
 \begin{equation} 
 \Delta_F(h,h) = - i \parallel h \parallel^2 \;. \label{hres}
 \end{equation} 
 Collecting everything, it turns out that 
 
 \begin{equation} 
\langle e^{i \varphi(h)}\; e^{i \varphi(h')} \rangle_{Feyn} = e^{\frac{i}{2} \Delta_F(h+h')} = e^{-\frac{\parallel h+h'\parallel^2}{2}}  \;, \label{feynmr}
\end{equation} 
 which is exactly the result obtained from the canonical quantization, eq.\eqref{vA}. Finally, we get 
 \begin{equation} 
 \langle e^{i \varphi(h)}\; e^{i \varphi(h')} \rangle_{Feyn} =  \langle 0| e^{i \varphi(h)}\; e^{i \varphi(h')} |0\rangle_{can} \;, \label{canequivproof}
\end{equation} 
showing thus the equivalence between the Feynman path integral and the canonical quantization for the Bell-CHSH inequality.

\section{Conclusion}\label{conclus}
In this work we have pursued the study of the Bell-CHSH inequality in relativistic Quantum Field Theory by implementing its formulation within the Feynman path integral. Both canonical quantization and functional integral yield the same expression for the correlation function of Weyl operators. \\\\This feature relies on the observation that, by construction, the Bell-CHSH combination is compatible with the fundamental principle of relativistic causality, as required by demanding that Alice and Bob be space-like separated. Moreover, the localization of Alice anf Bob in space-time can be given a precise mathematical formulation by employing a suitable set of smoorh test functions with compact support,  which act alike localizers for the bounded operators entering the Bell-CHSH inequality, which turns out to be compatible with the time ordering $T$, a key property for the path integral formulation. \\\\We strenghten that 
the Feynman path integral formulation of the Bell-CHSH inequality opens the door to many applications such as: 
\begin{itemize}
\item treatment of interacting field theories by employing the usual dictionary of Feynman diagrams

\item study of the Bell-CHSH inequality in Abelian and non-Abelian gauge theories in a manifest BRST invariant setting, through the use of the Faddeev-Popov BRST invariant action. In this regard, we refer to \cite{Peruzzo:2022pwv}, where the BRST invariant formulation for the Wyl operators of Yang-Mills theories in presence of Higgs fields has been  outlined. 

\item Finally, the path integral formulation might enable us to estimate possible non-perturbative contributions to the Bell-CHSH inequality stemming from the existence of soliton sectors of the theory under investigation. 

\end{itemize}

\section*{Acknowledgements}
The authors would like to thank the Brazilian agencies CNPq and FAPERJ for financial support. This study was financed in part
by the Coordena{\c c}{\~a}o de Aperfei{\c c}oamento de Pessoal de N{\'\i}vel Superior--Brasil (CAPES) --Finance Code 001. S.P.~Sorella is a level $1$ CNPq researcher under the contract 301030/2019-7.

\end{document}